\documentclass[prb,aps,floats,twocolumn,superscriptaddress]{revtex4-1}
\usepackage[english]{babel}
\usepackage{amsmath}
\usepackage{amsfonts}
\usepackage{graphicx}
\usepackage{color}
\usepackage{lipsum}
\usepackage{bm}
\pdfoutput=1

\newcommand{\bs}{\boldsymbol}

\DeclareMathOperator{\sgn}{sgn}

\begin{document}
\title {Dynamical Shiba states by precessing magnetic moments in an s-wave superconductor}
\author{V. Kaladzhyan}
\affiliation{Institut de Physique Th\'eorique, CEA/Saclay,
Orme des Merisiers, 91190 Gif-sur-Yvette Cedex, France}
\affiliation{Laboratoire de Physique des Solides, CNRS, Univ. Paris-Sud, Universit\'e Paris-Saclay, 91405 Orsay Cedex, France}
\affiliation{Moscow Institute of Physics and Technology, Dolgoprudny 141700, Moscow region, Russia}

\author{S. Hoffman}
\affiliation{Department of Physics, University of Basel, Klingelbergstrasse 82, CH-4056 Basel, Switzerland}

\author{M. Trif}
\affiliation{Laboratoire de Physique des Solides, CNRS, Univ. Paris-Sud, Universit\'e Paris-Saclay, 91405 Orsay Cedex, France}
\date{\today}

\begin{abstract}

We study theoretically the dynamics of a Shiba state forming around precessing classical
spin in an s-wave superconductor. Utilizing a rotating wave description for the precessing magnetic
impurity, we find the resulting Shiba bound state quasi-energy and the spatial extension of the Shiba wavefunction. We show that such a precession pertains to dc charge and spin currents flowing through a normal STM tip tunnel coupled to the superconductor in the vicinity of the impurity. We calculate these currents and find that they strongly depend on the magnetic impurity precession frequency,  precession  angle, and  on the position of the Shiba energy level in the superconducting gap. The resulting charge current is found to be proportional to the difference between the electron and hole wavefunctions of the Shiba state, being a direct measure for such an asymmetry. By dynamically driving the impurity one can infer the spin dependence of the Shiba states in the absence of a spin-polarized STM tip.   

\end{abstract}

\maketitle

\section{Introduction}

Magnetic  impurities inserted in $s$-wave superconductors can bind in-gap, spin-polarized electrons, in the so-called localized Shiba states \cite{Balatsky2006}. While these states have been predicted more than four decades ago \cite{Yu1965,Shiba1968,Rusinov1969,Sakurai1970}, and experimentally detected before the turn of the century\cite{Yazdani1997}, recently they have been receiving lots of attention as they have been suggested as building blocks for a topological superconductor. Similar to topological insulators, these materials have the properties that they are insulating in the bulk, but can possess metallic surface states, depending if they are in the so-called trivial (no surface states) or non-trivial (with surface states) regime. The surface states of a topological superconductor are Majorana fermions\cite{Fu-Kane}, exotic particles that are their own antiparticles, and which have been promoted as building blocks for a topological quantum computer.  In condensed matter systems they emerge as excitations in $p$-wave superconductors (at the edges of the sample, or bound to vortices) which, however, rarely exist in a natural form. Instead, various implementations have been put forward, all which rely on engineering  such new states of matter. One of the most viable ideas is to utilize magnetic impurities in conventional $s$-wave superconductors\cite{Choy2011,Nakosai2013,Yazdani2013,Braunecker2013,Klinovaja2013,Vazifeh2013,Pientka2013,Pientka2014,Poyhonen2014,Heimes2014,Reis2014,Ojanen2015,Peng2015,Rontynen2015,Braunecker2015,Zhang2016,Hoffman2016,Neupert2016,Kimme2016}, that give rise to Shiba states which, if put in close proximity to each other can form an electronic band that precisely shows the properties of a topological superconductor. 

There have been a lot of theoretical\cite{Brydon2015,Meng2015,Lutchyn2015} and, most importantly, experimental progress in this direction: single Shiba states have been visualized, both in 3D and 2D superconductors\cite{Shuai-Hua2008,Franke2015,Menard2015}, as well as a one-dimensional chain of Shiba impurities that gives rise to a Shiba band. \cite{Pientka2013,Pientka2014} Probing by the aid of STM techniques allows to reveal the spectral properties of these systems locally.  Moreover, it was showed experimentally that such a chain supports  zero energy states at its edge which were claimed to be  Majorana fermions\cite{Nadj-Perge2014,Ruby2015a}. Two-dimensional structures have been also addressed, both theoretically and experimentally\cite{Menard2016}, where magnetic impurities formed a two-dimensional Shiba band showing zero energy Majorana running modes at its egdes.

\begin{figure}[t] 
\centering
\includegraphics[width=0.85\linewidth]{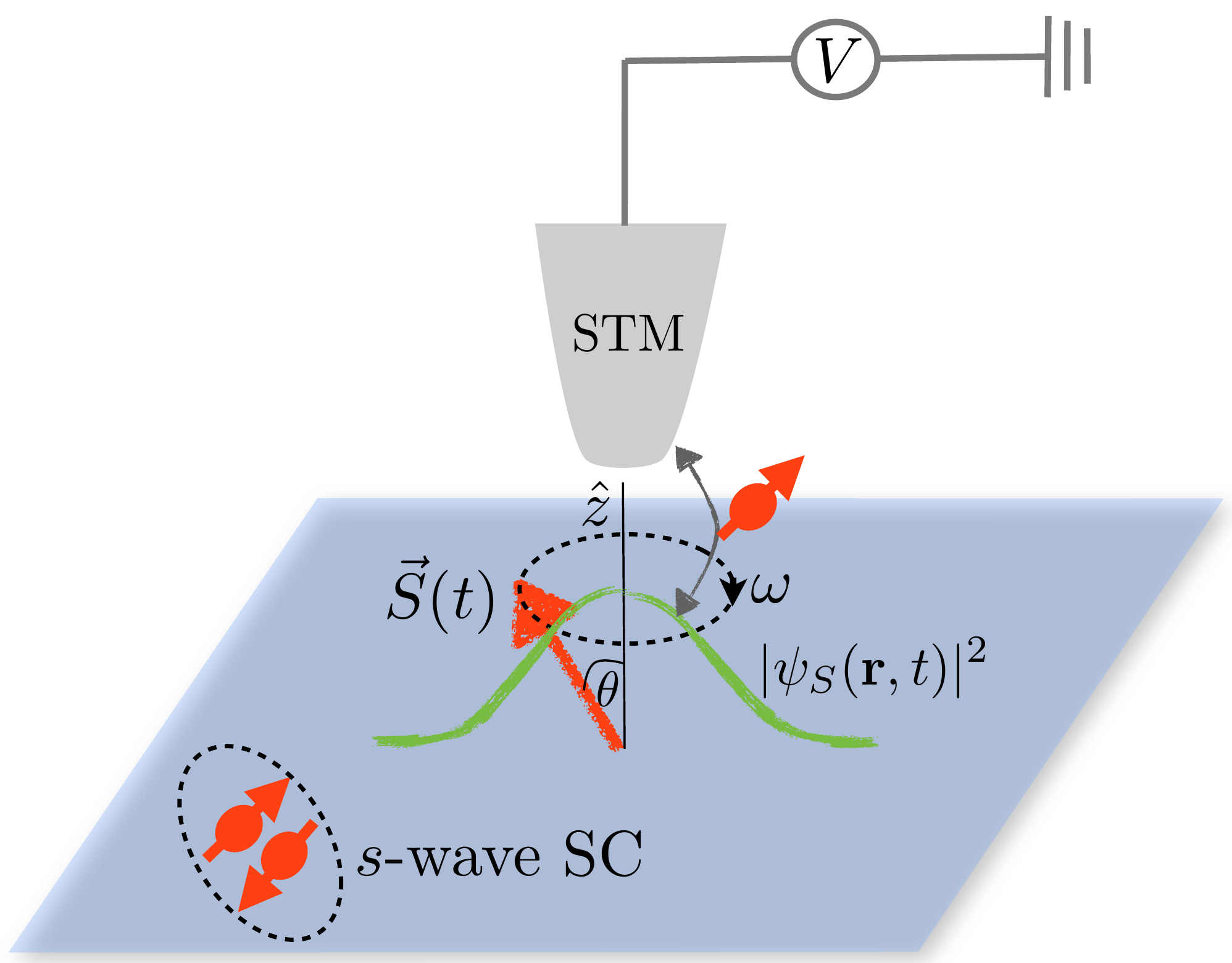}
\caption{The sketch of the dynamical Shiba setup. A classical spin (long red arrow) embedded in a two-dimensional superconductor (blue plane)  precesses at a frequency $2\omega$ and with an angle  $\theta$ with respect to the $z$ axis. A localized Shiba state forms around the classic spin, whose wave function $\psi_{S}(\bm{r},t)$ depends on both time and position. An STM tip couples to this localized state and detects a voltage due to the spin and charge currents pumped via the dynamical Shiba state.} 
\label{sketch_shiba} 
\end{figure}

Such studies did not address the out-of-equilibrium physics associated with the dynamics of the magnetic impurities. Magnetization dynamics is at the core of various effects in ferromagnets, such as, for example, the spin pumping and the spin-transfer torque, which allows to create and transfer the spin degree of freedom between different systems. It allows to visualize both the equilibrium, but more importantly, the out-of-equilibrium properties of a magnetic system. It is thus of crucial importance to analyze the interplay of magnetization dynamics and the Shiba physics in superconductors.

In this paper we study the magnetization dynamics of a single magnetic impurity in an $s$-wave superconductor, the buiding block of a Shiba impurity band. The precession of the impurity gives rise to a dynamical Shiba state that drives both an electrical and spin current through an STM tip in its proximity. We calculate these currents, and show that such a method allows to spectroscopically address the Shiba impurity and its dynamics.   	
 
The paper is organized as follows: In Sec.~\ref{Model} we introduce the setup and the model Hamiltonian of the dynamical Shiba states, in the presence of the STM tip. In Sec.~\ref{RWA} we develop the rotating wave approximation (RWA) of the model and describe the resulting dynamical Shiba states. Here we also investigate the (quasi-)energy spectrum and the time-dependent wave functions. In Sec.~\ref{transport} we switch on the coupling to the STM tip and analyze the charge and spin transport pertaining to the precessing magnetic moment through a normal STM tip. We analyze both the closed circuit (finite current, no voltage), as well as the open circuit (zero current, finite voltage) circuits.  Finally, in Sec.~\ref{conc} we end up with conclusions and some outlook on possible extensions for a chain of Shiba impurities.

\section{System and Model Hamiltonian}
\label{Model}

In this section we introduce the system and the model Hamiltonian. In Fig.~\ref{sketch_shiba} we show a sketch of the system: a precessing classical magnetic impurity in an $s$-wave superconductor. The precession could be induced, for example, by utilizing a microwave field from nearby stripline resonator that drives the impurity with a given frequency via the ferromagnetic resonance. The electrons in the superconductor are coupled to the impurity via the exchange coupling,  which  results in the occurrence  of an in-gap  Shiba state. An STM tip nearby the impurity is assumed to probe the emergent Shiba state via the out-of-equilibrium currents, both charge and spin traversing the tip due to the precessing impurity. The tip is assumed to be a normal metal, but generalizations to a ferromagnetic, or even superconducting tips are straightforward. 

The Hamiltonian describing the superconductor in the presence of the magnetic impurity, the STM tip, and its coupling to the superconductor, respectively, is given as:
\begin{align}
H_{tot}&=H_{S}+H_{M}+H_{T}\,,\nonumber\\
H_{S}&=\int d\bm{r}\psi^\dagger_S(\bm{r})[\epsilon_S\tau_z+\Delta_s\tau_x+J\delta(\bm{r})\bm{S}(t)\cdot\bm{\sigma}]\psi_S(\bm{r})\,,\nonumber\\
H_{M}&=\int d\bm{r}\sum_\sigma\psi^\dagger_{T\sigma}(\bm{r})\epsilon_T\psi_{T\sigma}(\bm{r}),\nonumber\\
H_{T}&=\mathcal{T}\sum_\sigma\left[\psi^\dagger_{S\sigma}(0)\psi_{T\sigma}(0)+\psi^\dagger_{T\sigma}(0)\psi_{S\sigma}(0)\right]\,,
\label{HamTot}
\end{align}
where $\psi^\dagger_S\equiv(\psi_{S\uparrow}^\dagger,\psi^\dagger_{S\downarrow},\psi_{S\downarrow},-\psi_{S\uparrow})$,  $\epsilon_{S(T)}=-\hbar^2\nabla^2/2m-\mu_{S(T)}$,  with $\mu_{S(T)}$ the chemical potential in the superconductor (STM tip), $\Delta_s$ is  the $s$-wave pairing  in the superconductor, $J$ is the exchange coupling between the classical spin $\bm{S}(t)=S_0(\sin{\theta}\sin{2b t},\sin{\theta}\cos{2b t},\cos{\theta})$ and the electrons in the superconductor, with $S_0$ and $\theta$ being the magnitude and the angle between the $z$ axis and the classical spin $\bm{S}$, respectively. Also, $\mathcal{T}$ is the hoping strength between the STM tip and the superconductor, $2b$ the magnetic impurity precession frequency, and  $\psi_{S\sigma}(\bm{r})$ [$\psi_{T\sigma}(\bm{r})$] and $\psi_{S\sigma}^\dagger(\bm{r})$ [$\psi_{T\sigma}^\dagger(\bm{r})$] are the annihilation (creation) operators for electrons at position $\bm{r}$ and spin $\sigma$ in the superconductor  and the STM tip, respectively.  Note that $\bm{\tau}=(\tau_x,\tau_y,\tau_z)$ and $\bm{\sigma}=(\sigma_x,\sigma_y,\sigma_z)$ are Pauli matrices that act in the Nambu and spin space, respectively. We mention that in general the magnetic impurity generates also a scalar potential, besides the magnetic exchange coupling. For the time being and for the clarity of the description, we disregard such a contribution, and we will discuss its effects (which turn out to be crucial) later in the section about the electronic transport.      

In the following, we analyze the above Hamiltonian in the absence of the coupling to the STM tip, thus only considering the precessing impurity.

\section{Rotating wave description of the superconductor}
\label{RWA}

The solution for the static Shiba impurity is well known. There are many works that describe the {\it static} Shiba impurity state, both their spectrum and wave function in various dimensions \cite{Yu1965,Shiba1968,Rusinov1969,Sakurai1970,Kaladzhyan2016}. Here we generalize those works for the case of the precessing impurity, and provide the theoretical framework for the experimental detection of such precession, and eventually of the Shiba state itself.  For simplicity, we assumed the impurity precesses circularly, otherwise the problem becomes more involved, though still tractable. There are various ways to approach the time-dependent problem. For periodic driving, like it is the case here, one approach is to utilize the so-called Floquet formalism, that is analogue to the Bloch wave function description of electrons in a periodic potential. The second approach is to utilize the rotating wave (RW) description, which implies switching to a rotating frame turning with the impurity, in which case the effective Hamiltonian becomes static, and perform calculations similarly to the static Shiba impurity. In the following we utilize the latter approach, well-suited when there are no spin-orbit interactions present.

 For that, we perform the following time-dependent transformation of the total Hamiltonian:
\begin{equation}
\tilde{H}_{tot}=U^\dagger(t) H_{tot}(t)U(t)-i\partial_t U^\dagger(t)U(t)\,,
\end{equation}
where we choose $U(t)=\exp{(i\sigma_z b t)}$. This transformation pertains to the following changes in Eq.~\eqref{HamTot}:  	
\begin{equation}
\bm{S}(t)\rightarrow\bm{S}(0)=S_0(\sin{\theta}\sin{\Phi},\sin{\theta}\cos{\Phi},\cos{\theta})\,,\\
\label{Spin}
\end{equation}
\begin{equation}
H_{j}\rightarrow H_{j}-b\int d\bm{r}\psi^\dagger_j(\bm{r})\sigma_z\psi_{j}(\bm{r})\,,
\label{RWAHam}
\end{equation} 
where $j=S,T$ and $\Phi\in[0,2\pi)$, while all the other terms stay the same under this transformation. In a nutshell, in the rotating frame the precessing spin becomes static and points along a direction defined in Eq.~\eqref{Spin}, while both the superconductor and the tip are subject to a fictitious magnetic field of size $b$ along the $z$ direction. Note that this fictitious field does not affect the superconductor gap, as it does not enter the gap equation. The wave functions in the two frames are related by $\psi(\bm{r},t)=U(t)\tilde{\psi}(\bm{r},t)$, with $\psi(\bm{r},t)$ [$\tilde{\psi}(\bm{r},t)$] being the wave function in the lab (rotating) frame.

Next we solve the spectrum of the Shiba state in the rotating frame. A natural way to proceed is by calculating the retarded Green's function of the superconductor in the presence of the impurity, and extracting the Shiba energy from its poles \cite{Balatsky2006}. This can be found from the Dyson equation: 
\begin{equation}
G_{S}^{-1}(\omega)=G_{0}^{-1}(\omega)-J\bm{S}(0)\cdot\bm{\sigma}\,,
\label{GreenFull}
\end{equation}   
where $G_0(\omega)$ is the bare Green function of the superconductor for frequency $\omega$ and at $\bm{r}=0$ in the presence of the fictitious magnetic field $b$, and which reads:
\begin{align}
G_0(\omega)&=-\frac{\pi\nu_0}{2}\sum_{\sigma=\pm1}\frac{\omega+\sigma b+\Delta_s\tau_x}{\sqrt{\Delta_s^2-(\omega+\sigma b)^2}}(1+\sigma\sigma_z)\,.
\end{align}  
Here, $\nu_0$ is the density of states of the superconductor in the normal state at the Fermi level. Note that in the limit $b\rightarrow0$ we recover the usual Green's function of a superconductor at position $\bm{r}=0$ \cite{Balatsky2006}. While we can now invert Eq.~\eqref{GreenFull} and find its poles, the general solution is too long to be displayed and it is not very insightful either. There are, however, a few limits worth depicting, in particular in the case  $\delta\equiv b/\Delta_s\ll1$. Such a limit also implies that  the bulk gap is not closed in rotating frame. Full analytical results are obtainable for two specific angles, $\theta=0$  and $\theta=\pi/2$. For $\theta=0$, we simply obtain $E_S(b)=\epsilon_S+b$, with $\epsilon_S=\Delta_s(1-\alpha^2)/(1+\alpha^2)$ being the Shiba state energy at $b=0$ \cite{Balatsky2006}, with $\alpha\equiv\pi JS_0\nu_0$, while for $\theta=\pi/2$ we get: 
\begin{align}
E_S(b)=\frac{(1+\alpha^4)-\sqrt{(1-\alpha^4)^2\delta^2+4\alpha^4}}{1-\alpha^4}\Delta_s\,,
\end{align}
which highlights the non-linear behavior of the Shiba energy as a function of the driving $b$. This becomes exactly $E_S(0)$ for $\alpha\rightarrow1$, which implies no dependence on the driving frequency. Moreover, we can infer the full $\theta$ dependence of the Shiba energy for the case $\alpha=1$:
\begin{equation}
E_S(b,\theta)=\sqrt{\frac{2\delta^2-\tan^2{\theta}+\tan^2{\theta}\sqrt{1-4\delta^2\cos^2{\theta}}}{2}}\Delta_s\,.
\end{equation}

Instead of utilizing exact solutions (simple in a few cases), it is instructive to extract approximate Shiba energies valid in the limit $\delta\ll1$,  and $\alpha\approx 1$ (the so-called deep Shiba limit). Then, as showed in Appendix~\ref{app1}, the Shiba states Green's function becomes: 
\begin{align}
G_S(\omega)\approx\frac{1}{\omega-E_S(b,\theta)}[1+(u^2-v^2)\tau_z+uv\tau_x](1+\sigma_\parallel)\,,
\end{align} 
with $E_S(b,\theta)\simeq \epsilon_S+b\cos{\theta}$ being the Shiba energy in the rotating frame. The correction to the energy of the Shiba state is of the order $\delta^2$, while the neglected entries of the matrix $G_S(\omega)$ are of the order $\delta\sin{\theta}/\alpha$, and they also lead to more complicated spin-mixing terms. In Appendix \ref{app1} we present the leading order corrections to the matrix $G_S(\omega)$.  Here $u$ and $v$ are the electron and hole components, respectively, of the Shiba state wave function, and $\sigma_\parallel=\bm{S}\cdot\bm{\sigma}/|\bm{S}|$, i.e. the spin projection along the magnetic impurity. In the case of purely magnetic scattering $u=v=1/\sqrt{2}$, but in general $u\neq v$  if scalar scattering is taken into account  \cite{Ruby2015b}. We will keep their relative value arbitrary in the following, and for their general expressions we refer the reader to Ref.~[\onlinecite{Ruby2015a}].   We mention that  the particle-hole symmetric solution $-E_{S}(b)$ can be found by applying the particle-hole conjugation operator $\mathcal{P}=\tau_y\sigma_yK$ to the above Shiba Green's function: $G_S(\omega)\rightarrow\mathcal{P}G_S(\omega)\mathcal{P}^\dagger$.

Physically, the above solution corresponds to a magnetic impurity in a superconductor and subject to an external magnetic field $B=b\cos{\theta}$ along the magnetic impurity direction. In the $\delta\ll1$ limit the transverse component of the fictitious magnetic field is suppressed and can be neglected, much in the same way such components are ineffective in a usual ferromagnet \cite{TserkovnyakPRB08}. However, while there the scale for neglecting these terms is the exchange field $J$, here the situation is more complicated, as it requires that both  $\delta\ll1$ and $\alpha\approx 1$. 
A few more comments are in order at this point.  For $b>\Delta_s$ the superconducting gap is closed in the rotating frame and the quasi-stationary Shiba state will always be resonant with the the continuum of states, and thus not protected. Even more, in the limit $\delta\gg1$, we recover the case of an impurity in a normal metal, and the associated effects, for example that in such a case there is no charge current flowing through the normal STM tip tunnel coupled to the impurity\cite{MoriyamaPRL08,TserkovnyakPRB08}.  On the other hand, for $\delta<1$, the superconducting gap in the rotating frame diminishes, i.e. it becomes $\Delta_s^{\rm eff}=\Delta_s-b$, and  Shiba states can occur within this gap. In the present work we focus only on the limit $\delta\ll1$, and thus assume such solutions are always present. 

For completeness, we also discuss briefly the position-dependent  wave-function of the driven Shiba state and, for simplicity, we focus on a two-dimensional superconductor.  At $\bm{r}=0$, this is given by $\tilde{\phi}_S(0)=(u,0,v,0)^T$. From this, we can readily find the $\bm{r}$ dependence of the Shiba wave-functions as follows \cite{Balatsky2006}: 
\begin{align}
\tilde{\phi}_S(\bm{r})=G_{0}(E_S,\bm{r})V_{\rm eff}\tilde{\phi}_S(0)\,,
\end{align}
where 
\begin{align}
G_{0}(E_S,\bm{r})&=\sum_{\sigma=\pm}\left[(E_S+\sigma b)X_0^\sigma(\bm{r})+X_1^\sigma(\bm{r})\right.\nonumber\\
&\left.+\Delta_sX_0^\sigma(\bm{r})\tau_x\right](1+\sigma\sigma_\parallel)\,,
\end{align}
with:
\begin{align}
X_0^{\sigma}(\bm{r})&=-\int\frac{d\bm{p}}{(2\pi)^2}\frac{e^{i\bm{p}\cdot\bm{r}}}{\xi_p^2+\Delta_s^2-(E_S+\sigma b)^2}\nonumber\\
&=-2\nu_0\frac{1}{\omega_\sigma}{\rm Im}\, \mathrm{K}_0\left[-i\left(1+i\frac{\omega_\sigma}{v_Fp_F}\right)p_Fr \right]\,,\\
X_1^{\sigma}(\bm{r})&=-\int\frac{d\bm{p}}{(2\pi)^2}\frac{\xi_p e^{i\bm{p}\cdot\bm{r}}}{\xi_p^2+\Delta_s^2-(E_S+\sigma b)^2}\nonumber\\
&=-2\nu_0{\rm Re}\,\mathrm{K}_0\left[-i\left(1+i\frac{\omega_\sigma}{v_Fp_F}\right)p_Fr\right]\,.
\end{align}
Above,  $K_0(x)$ is the Bessel function of second kind, and $\omega^2_\sigma \equiv \Delta_s^2-(E_S+\sigma b)^2$. Detailed derivation of these expressions is left for the Appendix \ref{app2}.  These expressions allow to extract the full wave function of the dynamical Shiba states, or to perform the  small expansion $b$ in order to extract approximate expressions for the electron and hole components at a distance $\bm{r}$ from the impurity, which in turn would allow, as described in the next section, to calculate currents detected at that distance.




\section{Charge and spin currents}
\label{transport}

In the previous section we calculated the spectral properties of the dynamical Shiba state forming around a precessing magnetic moment. Here we address the electronic transport between the dynamically formed Shiba state and an STM tip brought into its proximity. We will derive both the charge and spin currents flowing through the STM tip due to the magnetic precession at $\bm{r}=0$. The generalization $\bm{r}\neq0$ is straightforward by utilizing the results in the previous section. We will consider two cases: the closed and open circuits, which in turn will give a finite current in the absence of a voltage and a finite voltage in the absence of a current, respectively. 

The charge and spin current operators are defined as follows:  
\begin{align}
\hat{I}_{c}&\equiv ie[H_T,N_T]=ie\mathcal{T}\sum_{\sigma}\left[\psi_{S\sigma}^\dagger(0)\psi_{T\sigma}(0)-{\rm H. c. }\right]\,,\\
\hat{\bm{I}}_{s}&\equiv i\hbar[H_T,\bm{S}_M]=i\hbar \mathcal{T}\sum_{\sigma}\left[\psi_{S\sigma}^\dagger(0)(\bm{\sigma})_{\sigma\sigma'}\psi_{T\sigma'}(0)-{\rm H. c. }\right]\,,
\end{align}
where we utilized the following substitutions: $N_T=\sum_{k,\sigma}c^\dagger_{Tk\sigma}c_{Tk\sigma}$ for the number of electrons, and
\begin{align}
\bm{S}_T(0)&=\hbar\sum_{k,\sigma}c^\dagger_{Mk\sigma}(\bm{\sigma})_{\sigma\sigma'}c_{Mk\sigma'}\,,
\end{align}
for the total number of spins in the  STM tip, respectively.  In order to calculate the average current, we utilize the non-equilibrium Green's function technique, and for details on the calculations we refer the reader to the  Refs.~[\onlinecite{Ruby2015b}] and [\onlinecite{Martin2014}]. The expression for the charge current reads:
\begin{align}
I_{c}(t)&=\frac{2e\mathcal{T}}{\hbar}{\rm Tr}\left[\tau_z\left(\tau_z\hat{G}^<_{ST}(t,t)-\hat{G}^<_{TS}(t,t)\tau_z\right)\right]\,,
\end{align} 
while the spin current $\bm{I}_s(t)$ is found by substituting the first $\tau_z$ with $\bm{\sigma}$. Above, $\hat{G}^<_{ij}(t',t)=i\langle\psi_i^{\dagger}(t)\psi_j(t')\rangle$, is the lesser Green's function in the Nambu $\otimes$ spin space ($4\times4$ matrix), with $i,j=S,T$. Note that $(\psi_i)^T\equiv(\psi^\dagger_{i\uparrow},\psi^\dagger_{i\downarrow},\psi_{i\downarrow},-\psi_{i\uparrow})$.  We are thus left with the task of calculating the $\hat{G}^<_{ij}(t',t)$ at $t=t'$. While this is a time-dependent problem, by switching to the rotating frame, we can proceed to calculate the transport quantities as for the time-independent case. In the frequency domain, the following relations hold \cite{jauhoBook,Ruby2015a}:  
\begin{align}
G_{ST}^{<}(\omega)&=\mathcal{T}\left[G_S^r(\omega)\tau_zg_T^<(\omega)+G_S^<(\omega)\tau_z g_T^a(\omega)\right]\,,\\
G_{TS}^{<}(\omega)&=\mathcal{T}\left[g_T^r(\omega)\tau_zG_S^<(\omega)+g_T^<(\omega)\tau_z G_S^a(\omega)\right]\,,
\label{Langreth}
\end{align}
where $G_S^{<,r,a}(\omega)$ $\left( g_T^{<,r,a}(\omega) \right)$ are the lesser, retarded and advanced Green's functions of the superconductor (STM tip). Using these relations, and rearranging the terms accordingly, we obtain the following well-known expression for the current \cite{jauhoBook}:
\begin{align}
\!\!I_{c}&=\frac{2e\mathcal{T}^2}{\hbar}\int d\omega{\rm Tr}\left[G_S^>(\omega)g_T^<(\omega)-G_S^<(\omega)g_T^>(\omega)\right] \,.
\end{align}

\begin{figure}[t] 
\centering
\includegraphics[width=0.99\linewidth]{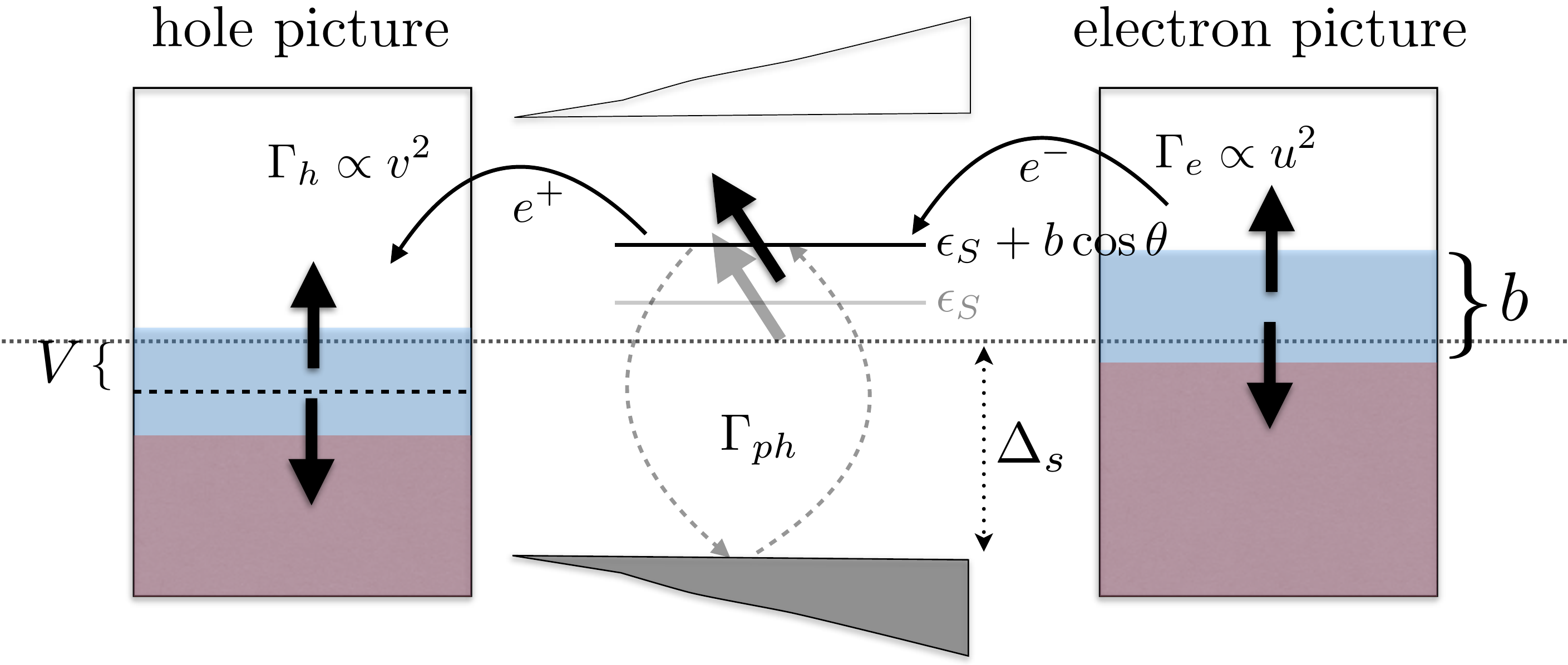}
\caption{A sketch of the transport channels in the rotating frame, in the presence of spin ($b$) and voltage  ($V$) biases. Both the electron and hole pictures are presented, with the corresponding rates $\Gamma_{e(h)}\propto u^2(v^2)$ (see text for details). The original Shiba state $\epsilon_S$ and the one in the rotating frame $E_S(b,\theta)=\epsilon_{S}+b\cos{\theta}$ are shown. The Shiba state is thermalized via internal relaxation processes that are accounted for by the rate $\Gamma_{ph}$ that mediates transitions between the bulk and the Shiba state.} 
\label{SketchTransport} 
\end{figure}

Thus, we need only the lesser and greater Green's functions of the superconductor and the tip respectively. Since we are interested in the transport through the in-gap Shiba state, we will utilize only the Green's function of this state, and not the whole superconductor. Moreover, for the transport to occur, we need to account for various processes that give rise to a finite lifetime of the Shiba state, such as photon, or phonon-assisted relaxation (or excitation) into the gaped continuum of states, and due to the coupling to the tip\cite{Ruby2015b}. We get (for details regarding the derivation see Appendix~\ref{app3}):
\begin{align}
G_S^<(\omega)&=i\frac{n_S\Gamma_{ph}+\sum_{\sigma,\tau}\Gamma_\tau n_{\tau,\sigma}(1+\sigma\cos{\theta})}{[\omega-E_S(b,\theta)]^2+(\Gamma_{\rm tot}/2)^2}\nonumber\\
&\times [1+(u^2-v^2)\tau_z+uv\tau_x](1+\sigma_\parallel)\,,
\end{align} 
where $\Gamma_{\rm tot}=\Gamma_e+\Gamma_h+\Gamma_{ph}$, with  $\Gamma_{e(h)}=2\pi\nu_0\mathcal{T}^2u^2(v^2)$, $\Gamma_{ph}$ is the intrinsic width of the Shiba state (due to photons or phonons, for example), $n_{\tau\sigma}(\omega)=n_F(\omega-\tau V-\sigma b)$, with $\tau,\sigma=\pm1$ are the Fermi distribution functions in the STM tip for electrons ($\tau=1$) and holes ($\tau=-1$) with a given spin $\sigma$,  while  $V$ is the applied, or the induced voltage in the combined system. Moreover, $n_S=n_F[\omega-E_S(b,\theta)]$. We mention that  $G_S^>(\omega)$ can be found simply by substituting all Fermi functions by $n_F\rightarrow-1+n_F$ in the above expression. Note that the spin direction quantization is different in the two systems: the Shiba state is defined along the magnetization $\bm{m}(0)$, while the fictitious biases in the STM tip are defined along the $z$ direction. The $\alpha=a,r,>,<$ Green's function of the STM tip in the Shiba spin basis can be written as $g_T^\alpha(\omega)=U(\theta)\tilde{g}_{T}^\alpha U^{\dagger}(\theta)$, with $U(\theta)=\exp{(i\theta\sigma_y/2)}$, and
\begin{equation}
\tilde{g}^\alpha_T(\omega)=\sum_{\sigma}\tilde{\rm g}_{T,\sigma}^\alpha(\omega+\sigma b+\tau\tau_z)(1+\tau\tau_z)(1+\sigma\sigma_z)\,,
\label{NormGreen}
\end{equation}
being the Green's function of the STM tip in the $z$ spin basis, with $\tilde{\rm g}_{T,\sigma}^\alpha(\omega)$ the components of the Green's function of a bare Fermi gas at position $\bm{r}=0$. Putting everything together in the expression for the current, we obtain $I_c=I_c^s+I_c^a$, with
\begin{align}
I_c^s&=\frac{e}{h}\int d\omega\sum_{\sigma,\tau}\frac{\tau\Gamma_{ph}\Gamma_\tau(n_{\tau\sigma}-n_S)(1+\sigma\cos{\theta})}{[\omega-E_S(b,\theta)]^2+(\Gamma_{\rm tot}/2)^2}\,,\\
I_c^a&=\frac{2e}{h}\int d\omega\sum_{\sigma,\tau}\frac{\tau\Gamma_e\Gamma_h n_{\tau\sigma}(1+\sigma\cos{\theta})}{[\omega-E_S(b,\theta)]^2+(\Gamma_{\rm tot}/2)^2}\,,
\label{tot_current}
\end{align}
which describe the single particle and the Andreev reflection induced charge currents respectively. In Fig.~\ref{SketchTransport} we show a sketch of the transport processes giving rise to a current, both in the electron and hole pictures, along with the possible relaxation channels. For completeness, we assume the presence of both a spin and voltage biases driving out-of-equilibrium currents. The latter arises  in the case when the circuit formed by the Shiba state and the STM tip is open, i.e there is no current flowing, but a finite  voltage bias drop builds across the tunneling region. In the closed circuit instead, the voltage is zero, but there is a dc charge current flowing. In this case, $n_{\tau\sigma}\equiv n_\sigma$, which implies $I_c^a\equiv0$ and $I_c^s\propto(\Gamma_{e}-\Gamma_h)$. This means that for a current to flow in the presence of magnetization precession the electron and hole components of the Shiba state must be different. Such a difference arises naturally if one assumes, besides the spin scattering, the presence of a scalar potential too, as mentioned already in the previous section. This is in contrast to the usual voltage probed Shiba state, where such an asymmetry is not required, and the current is proportional to  $u^2$ ($v^2$) for voltages $V=\epsilon_s$ ($V=-\epsilon_s$).   

\begin{figure}[t] 
\centering
\includegraphics[width=0.99\linewidth]{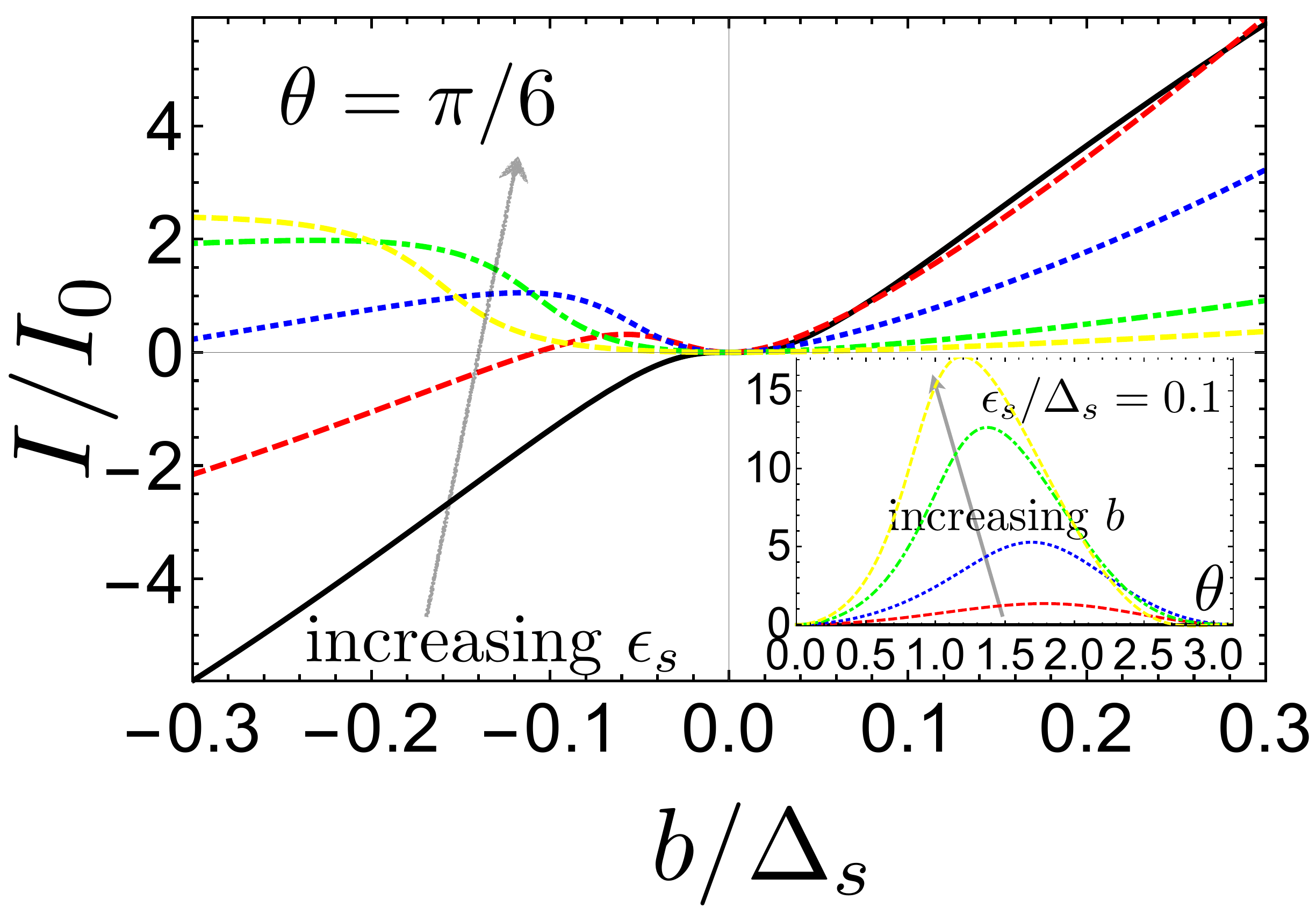}
\caption{Main: The magnetic impurity precession induced charge current in the closed circuit as a function of the precession frequency $b$ for various values of the Shiba energy $\epsilon_S$. Only the $I_c^s$ contributes to the current. The current is a measure of the presence of the Shiba impurity and highlights its position. For the plots we assumed $\epsilon_S=0.1,0.1,0.3$, and $0.4$. Inset: the current as a function of $\theta$ for $b=0,0.05,0.1,0.2$, and $0.3$.  All energies are expressed in terms of the superconducting gap $\Delta_s$, and $I_0=(e/\hbar\Delta_s)\Gamma_{ph}(\Gamma_e+\Gamma_h)$, while we assumed $(\Gamma_e-\Gamma_h)/(\Gamma_e+\Gamma_h)=0.6$} 
\label{ChargeCurrent} 
\end{figure}

In Fig.~\ref{ChargeCurrent} we show the dependence of the pumped dc current into the STM tip as a function of the precession frequency (main plot) and the precession angle (inset). In this case, there is no voltage bias present and the entire current is given by the single particle current. The current is a non-monotonic function of $b$, and shows peaks when it becomes similar to the bare Shiba energy, i. e. $\epsilon_S\simeq b\cos{\theta}$. The maximum voltage is found for $\theta\approx\pi/2$, which shifts to lower precession angles as the precession frequency is increased. We mention that the current has the symmetries $I_c(b,\theta)=I_c(-b,\pi-\theta)$. 
	
\begin{figure}[t] 
\centering
\includegraphics[width=0.99\linewidth]{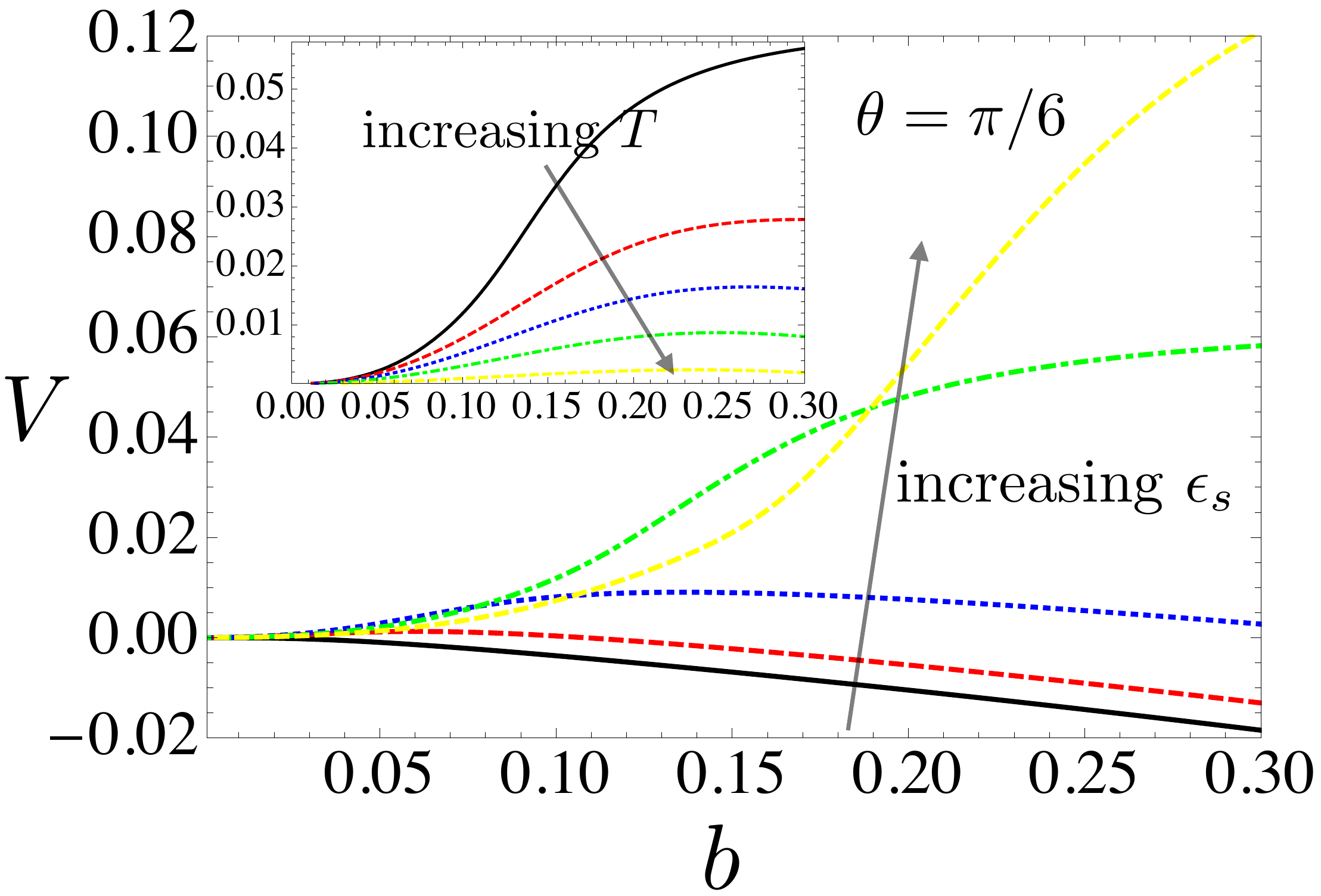}
\caption{Main: The magnetization precession induced voltage in the open circuit as a function of the precession frequency  $b$ for various values of the Shiba energy $\epsilon_S$. Only the $I_c^s$ contributes to the current. For the plots we assumed $\epsilon_S=0.1,0.1,0.3$, and $0.4$. Inset: the induced voltage as a function of $b$ for different temperatures,  $T=0,0.05,0.1,0.2$, and $0.3$.  All energies are expressed in terms of the superconducting gap $\Delta_s$, and we assumed $(\Gamma_e-\Gamma_h)/(\Gamma_e+\Gamma_h)=0.6$} 
\label{Voltage} 
\end{figure}

In the case of a static impurity that is voltage probed by an STM tip, one defines the differential conductance of the system $G=dI/dV$. Similarly, in the present case one can define a differential magnetoconductance as $G_b(b,\theta)=dI_c/db$, which at zero temperature acquires the following simple form:
\begin{align}
\hspace{-0.35cm}G_b&=\frac{4b\Gamma_{ph}(\Gamma_e-\Gamma_h)E_S(b,\theta)\sin^2{\theta}}{[b^2+E_S(b,\theta)^2+(\Gamma_{\rm tot}/2)^2]^2-4b^2E^2_S(b,\theta)}\,.
\label{DiffCond}
\end{align}
This vanishes in the static case $b=0$, as well as for  $b=\epsilon_S/\cos{\theta}$, and thus shows a pronounced non-monotonic behavior as a function of $b$ that can be utilized to map out the spectral properties of the Shiba state. As stated before, in the case of an open circuit one detects a voltage instead of a current. By imposing the $I_c=0$, we obtain the induced voltage in the circuit, which is depicted in the main plot in Fig.~\ref{Voltage}. More precisely, in the main plot we consider the voltage at $T=0$, while in the inset we depict the variation of one of the curves for various temperatures. Depending on the energy of the Shiba state, the induced voltage can be either positive or negative, or it can even be non-monotonic as a function of the precession frequency.  Note that relatively large values (of the order  $0.1b$) are achievable for such a voltage, and they are not spoiled by finite temperature effects as long as this is not larger than the Shiba energy $T\leq E_S(b,\theta)$.   

\begin{figure}[t] 
\centering
\includegraphics[width=0.99\linewidth]{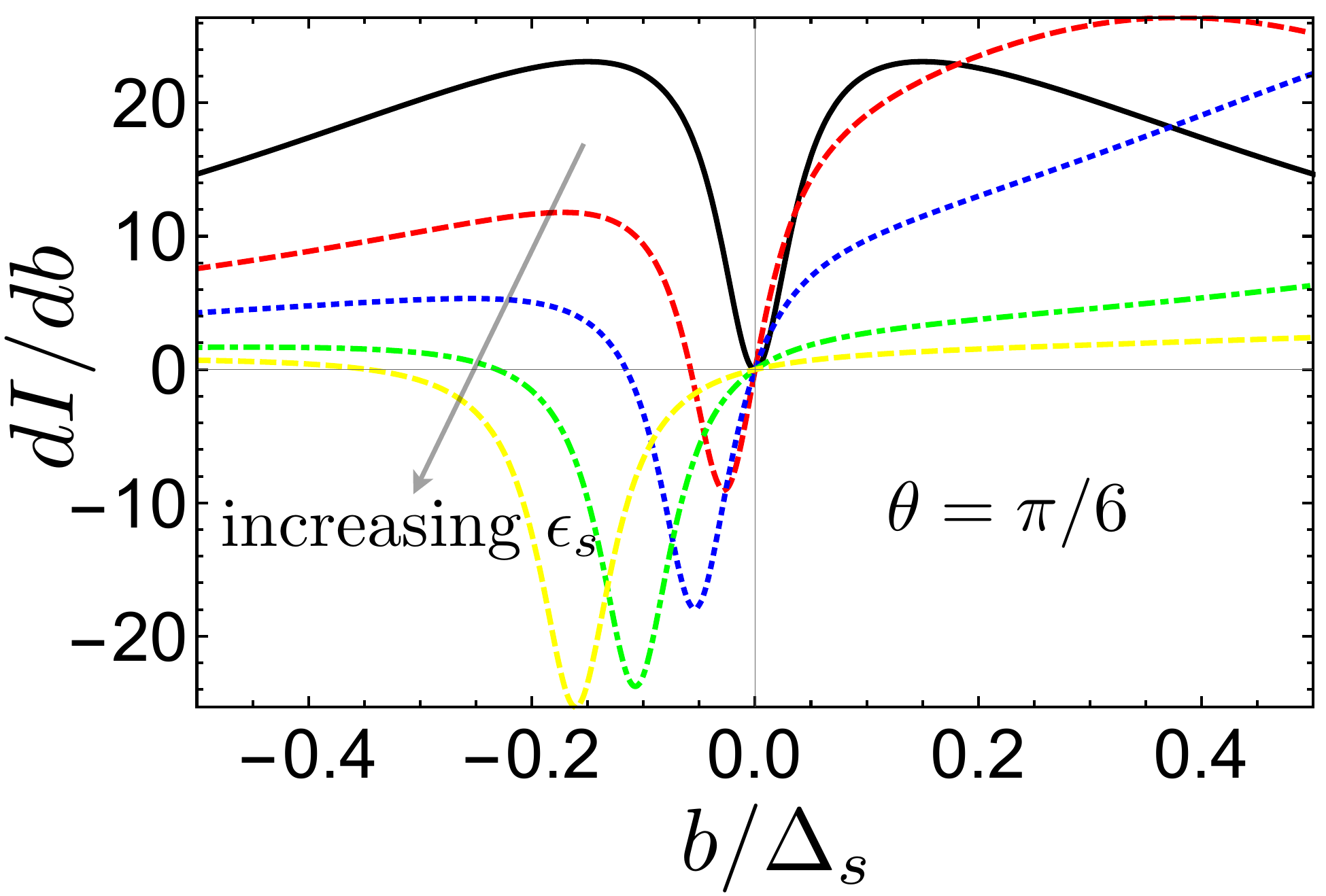}
\caption{Dependence of the differential magnetoconductance $G(b,\theta)$ in Eq.~\ref{DiffCond} on the precession frequency $b$ for $\epsilon_S=0$, $0.1$, $0.2$, $0,3$, and $0.4$, and for $\theta=\pi/6$. } 
\label{DiffField} 
\end{figure}

We can also evaluate the spin current pertaining to the magnetization dynamics, and we only assume the $z$ component. 
\begin{equation}
I_s^z=h\int d\omega\sum_{\sigma,\tau}\frac{\Gamma_{ph}\Gamma_\tau(n_{\tau\sigma}-n_S)(1+\sigma\cos{\theta})}{[\omega-E_S(b,\theta)]^2+(\Gamma_{\rm tot}/2)^2}\,,
\end{equation}  
which vanishes for vanishing $\theta$ or $b$. The efficiency of spin injection from the precessing magnetic impurity can be quantified by a spin magnetoconductivity, $G_s(b,\theta)=dI_z/db$, which differs from the charge counterpart only in the change $\Gamma_e-\Gamma_h\rightarrow\Gamma_e+\Gamma_h$. We thus refer to the previous results on the charge current with the aformentioned substitution. Note that for the spin current to flow, there is no need for particle-hole asymmetry in the Shiba wave function, in contrast to charge.

\section{Conclusions and Outlook}
\label{conc}

In this paper we studied the dynamics of a magnetic impurity in an $s$-wave superconductor, and its manifestations in the tunneling current through an STM tip. We found that in leading order in the precession frequency $b$ (as compared to the bulk superconductor gap $\Delta_s$), the main effect of the precession is to give rise to a shift of the Shiba levels by $b\cos{\theta}$, while the bulk gap is reduced as $\Delta^{\mathrm{eff}}_s(b)=\Delta_s-b$. We found that such a precession manifests itself in an current flow that can be detected by an STM tip. Utilizing the non-equilibrium Green's function technique, we calculated both the charge and spin currents flowing into the STM tip, and found that it gives a direct measure of the presence of the Shiba state. Moreover,  in analogy with the voltage biased systems, where one defines the differential conductance as measure of the presence of in-gap Shiba states, here we defined a differential magnetoconductance, $G_b=dI_c/db$, that reflects the resolution of the magnetization precession into a voltage bias. We found that this quantity reveals the Shiba state level, and that it vanishes for frequencies much larger than the superconducting gap. As opposed to static STM measurement, where the system is voltage biased, the asymmetry between the electron and hole components of the Shiba state is crucial: no current is flowing for a perfectly symmetric particle-hole Shiba state and this could be utilized to extract such an asymmetry, as well as to reveal the relevance of the spin degree of freedom.  

Our study was conducted assuming the precession of the magnetic moment was imposed externally by the magnetic fields of  a nearby microwave cavity. An extension of our theory could be done by accounting for the dynamics of the magnetic moment in a self-consistent manner, for example by utilizing the Landau-Lifshitz-Gilbert equation in the presence of the superconducting condensate. That, one hand, will describe the combined dynamics, while on the other hand, would allow one to extract  the influence of the Shiba state on the magnetization dynamics itself, which could be probed by the nearby microwave cavity. 

Finally, an extension of the time-dependent problem to a chain of Shiba impurities should allow to study the interplay between the magnetization dynamics and the emergence of in-gap topological superconductivity, or the so-called Shiba bands\cite{Pientka2013}. Moreover, it should be possible, by means of driving, to change the band structure, and even topology\cite{KaladzhyanTrifSimon}, similarly to the well-known Floquet topological insulators.

\acknowledgments

We acknowledge useful discussions with Marco Aprili, Tristan Cren, and Freek Masse. 

\appendix
\begin{widetext}

\section{Shiba state propagator in leading order in $\boldsymbol{\delta=b/\Delta_s}$}
\label{app1}

In this Appendix we present the leading order correction in $\delta$ for the propagator of the Shiba state. We start from the Dyson equation in the presence of impurity, and here we consider both a magnetic and scalar scattering, respectively. The scalar scattering Hamiltonian reads:
\begin{equation}
H_{V}=\int d\bm{r}\psi^\dagger_S(\bm{r})\tau_zV\delta(\bm{r})\psi_S(\bm{r})=\psi_S^\dagger(0)\tau_zV\psi_S(0)\,,
\end{equation} 
with $V$ the stength of the scalar scattering. Next we follow the derivation in Ref.~[\onlinecite{Ruby2015a}], adapted to our case of a Zeeman-split superconductor. The Dyson equation for the full Green's function reads:
\begin{equation}
G_{S}^{-1}(\omega)=G_{0}^{-1}(\omega)-J\bm{S}(0)\cdot\bm{\sigma}-V\tau_z\,,
\end{equation}
where $G_0(\omega)$ is the Green's function of the bulk superconductor in a Zeeman field,  depicted in Eq.~\eqref{GreenFull}, and  which can be solved in full glory, but here we present only the approximate expression for this function, along with the expressions for the Shiba energy and the electron and hole coherence factors. In order to connect with the case of zero driving ($b=0$), it is instructive to switch to a reference frame in which the static impurity (in the rotating frame) points along the $z$ axis. This means we need to perform a unitary rotation on the full superconducting Hamiltonian $H_{S}\rightarrow U(\theta)H_SU^\dagger(\theta)$, where $U(\theta)$ is given before Eq.~\eqref{NormGreen}. The general expression for the Shiba Green's function reads:
\begin{align}
G_S\approx\frac{1}{\omega-E_S(b,\theta)}\underbrace{\left(
\begin{array}{cccc}
u_\uparrow^2 & u_\uparrow u_\downarrow & u_\uparrow v_\downarrow & u_\uparrow v_\uparrow\\
u_\uparrow u_\downarrow & u_\downarrow^2 & u_\downarrow v_\downarrow & u_\downarrow v_\uparrow \\
u_\uparrow v_\downarrow & 	v_\downarrow u_\downarrow & v_\downarrow^2 & v_\downarrow v_\uparrow\\
v_\uparrow u_\uparrow & v_\uparrow u_\downarrow & v_\uparrow v_\downarrow & v_\uparrow^2 
\end{array}
\right)\,,
}_{\displaystyle{M_S}}
\end{align} 
where in leading order in  $\delta=b/\Delta_s$, the entries of the matrix read:
\begin{align}
u_\uparrow^2&=\frac{2\alpha\pi\nu_0\Delta_s\left[1+(\alpha+\beta)^2\right]}{\left[(1-\alpha^2+\beta^2)^2+4\alpha^2\right]^{3/2}}\\
v_\downarrow^2&=\frac{2\alpha\pi\nu_0\Delta_s\left[1+(\alpha-\beta)^2\right]}{\left[(1-\alpha^2+\beta^2)^2+4\alpha^2\right]^{3/2}}\\
u_\uparrow u_\downarrow&=v_\uparrow v_\downarrow=\frac{\delta\sin{\theta}}{4\alpha}\\
u_\uparrow v_\downarrow&=\frac{2\alpha\pi\nu_0\Delta_s}{\left[(1-\alpha^2+\beta^2)^2+4\alpha^2\right]^{2}}\\
u_\uparrow v_\uparrow&=-\frac{\delta[1+(\alpha-\beta)^2]\sin{\theta}}{\left[(1-\alpha^2+\beta^2)^2+4\alpha^2\right]^{2}}\\
u_\downarrow v_\downarrow&=-\frac{\delta[1+(\alpha+\beta)^2]\sin{\theta}}{\left[(1-\alpha^2+\beta^2)^2+4\alpha^2\right]^{2}}\,,
\end{align}
while all the other terms are of order $\mathcal{O}(\delta^2)$ and we neglect them in the following, i.e. we take $u_\downarrow^2=v_\uparrow^2=u_\downarrow v_\uparrow=0$. Here,  $\alpha=2\pi\nu_0J$ and $\beta=2\pi\nu_0V$, and we note that  the Shiba energy, in leading order in $b$ it is given by $E_S=\epsilon_S+b\cos{\theta}$, as already stated in the main text. We can see that if we take the zeroth order correction in $\delta$, that corresponds to the Shiba Green's function in Ref.~[\onlinecite{Ruby2015a}] with the energy $E_S$, and with $u_\uparrow\neq v_\downarrow$ for $\beta\neq0$. Note that for the terms $\propto\delta$ to be small, the condition $\alpha\approx 1$ needs to be met also, as stressed in the main text (i.e.  the system is in the deep Shiba limit). With all these approximations, we can rewrite the Shiba Green's function in the form shown in the main text, i.e.:
\begin{equation}
G_S(\omega)\approx\frac{1}{\omega-E_S(b,\theta)}\underbrace{\left[1+(u^2-v^2)\tau_z+uv\tau_x\right](1+\sigma_\parallel)}_{\displaystyle M_S^{\rm eff}}\,,
\end{equation}
where $u\equiv u_\uparrow$, and $v\equiv v_\downarrow$ in the above expression.  

\section{Green's function at finite $\bm{r}$ and the integrals}
\label{app2}

We write the Green's function in coordinate space in the following form:
\begin{align}
G_{0}(E_S,\bm{r})&=\sum_{\sigma=\pm}\left[(E_S+\sigma b)X_0^\sigma(\bm{r})+X_1^\sigma(\bm{r})+\Delta_sX_0^\sigma(\bm{r})\tau_x\right](1+\sigma\sigma_\parallel)\,.
\end{align}
To find the coordinate dependence we need to calculate the following integrals:
\begin{align}
X_0^\pm(\bs r) &= -\int \frac{d\bs p}{(2\pi)^2} \frac{e^{i \bs{pr}}}{\xi_p^2 + \Delta_s^2 - (E\pm b)^2}\,, \\
X_1^\pm(\bs r) &= -\int \frac{d\bs p}{(2\pi)^2} \frac{\xi_p\, e^{i \bs{pr}}}{\xi_p^2 + \Delta_s^2 - (E\pm b)^2}\,,
\end{align}
We linearize the spectrum around Fermi level $\xi_p = v_F (p-p_F)$ and rewrite $\int \frac{d\bs p}{(2\pi)^2} = \nu_0 \int d\xi_p$. Since we are computing the retarded Green's function we add an infinitesimal positive energy shift $E \to E + i0$:
\begin{align*}
X_0^\pm(\bs r) = -\int \negthickspace \frac{d\bs p}{(2\pi)^2} \frac{e^{i \bs{pr}}}{\xi_p^2 + \Delta_s^2 - (E\pm b)^2} = -\nu_0 \negthickspace \int\negthickspace d\xi_p \negthickspace \int \frac{d\phi_p}{2\pi} \frac{e^{i p r \cos \left(\phi_p - \phi_r \right)}}{\xi_p^2 + \Delta_s^2 - (E\pm b)^2 - i\sgn(E\pm b)\cdot 0} = \phantom{aaaa}\\
= -\nu_0 \left[\mathcal{P}\negthickspace\int \negthickspace d\xi_p\frac{J_0\left[\left(p_F + \xi_p/v_F \right) r \right]}{\xi_p^2 + \Delta_s^2 - (E\pm b)^2} + i\pi \sgn(E\pm b) \int \negthickspace d\xi_p \cdot \delta(\xi_p^2 + \Delta_s^2 - (E\pm b)^2)\cdot J_0\left[\left(p_F + \xi_p/v_F \right) r \right] \right]
\end{align*}
\begin{align*}
X_1^\pm(\bs r) = -\int \negthickspace \frac{d\bs p}{(2\pi)^2} \frac{\xi_p \, e^{i \bs{pr}}}{\xi_p^2 + \Delta_s^2 - (E\pm b)^2} = -\nu_0 \negthickspace \int \negthickspace d\xi_p \negthickspace \int \frac{d\phi_p}{2\pi} \frac{\xi_p\, e^{i p r \cos \left(\phi_p - \phi_r \right)}}{\xi_p^2 + \Delta_s^2 - (E\pm b)^2 - i\sgn(E\pm b)\cdot 0} = \phantom{aaaa}\\
= -\nu_0 \left[\mathcal{P}\negthickspace\int \negthickspace d\xi_p\frac{\xi_p J_0\left[\left(p_F + \xi_p/v_F \right) r \right]}{\xi_p^2 + \Delta_s^2 - (E\pm b)^2} + i\pi \sgn(E\pm b) \int \negthickspace d\xi_p \cdot \delta(\xi_p^2 + \Delta_s^2 - (E\pm B)^2)\cdot \xi_p J_0\left[\left(p_F + \xi_p/v_F \right) r \right] \right]\,.
\end{align*}
We assume the (effective) Shiba energy satisfies $E_S(b,\theta) \in \left( -\Delta_s+b, \Delta_s-b \right)$  and we denote  $\omega_\pm = \sqrt{\Delta_s^2 - (E\pm b)^2}$. We get: 
\begin{align*}
X_0^\pm(\bs r) &=-\nu_0\mathcal{P}\negthickspace\int \negthickspace d\xi_p\frac{J_0\left[\left(p_F + \xi_p/v_F \right) r \right]}{\xi_p^2 + \omega_\pm^2} = -2\nu\cdot \frac{1}{\omega_\pm} \mathrm{Im} \mathrm{K}_0 \left[ -i \left(1+i\Omega_\pm \right)p_F r \right]\,, \\
X_1^\pm(\bs r) &=-\nu_0\mathcal{P}\negthickspace\int \negthickspace d\xi_p\frac{\xi_p J_0\left[\left(p_F + \xi_p/v_F \right) r \right]}{\xi_p^2 + \omega_\pm^2} = -2\nu\cdot \mathrm{Re} \mathrm{K}_0 \left[ -i \left(1+i\Omega_\pm \right)p_F r \right]\,,
\end{align*}
where $\Omega_\pm \equiv \frac{\omega_\pm}{v_Fp_F}$. For further details see Ref.~[\onlinecite{Kaladzhyan2016}].

\section{Derivation of the Shiba Green function in the presence of the STM}
\label{app3}

Here we give more details on the derivation of the charge and spin currents flowing through the STM tip assuming the approximations above: $\delta\ll1$ and $\alpha\approx1$. In order to calculate the currents, we need to evaluate the lesser Green's functions $G_{RL}^<(t_1,t_2)$ and  $G_{LR}^<(t_1,t_2)$, as stated in the main text. While for a voltage biased Shiba state, one can utilize the simplified  $2\times2$ matrix description, here, because of the non-collinearity of the spin distributions in the superconductor (more precisely, of the in-gap Shiba state) and the STM tip, respectively, we have to consider the full $4\times4$ matrix in order to correctly describe the transport.  We assume, as in Ref.~[\onlinecite{Ruby2015a}], that there are two contributions to the Shiba Green's function self-energy: one intrinsic, due to photon-assisted excitations between the Shiba state and the bulk, and one due to the coupling to the tip. The first contribution can be taken into account as follows \cite{Ruby2015a}:
\begin{align}
g_{S}^{r,a}(\omega)&=\frac{1}{\omega-E_S(b,\theta)\pm i\Gamma_{ph}/2}M_S^{\rm eff}\,,\\
g_{S}^{<,>}(\omega)&=\frac{\Sigma^{<,>}_{ph}(\omega)}{[\omega-E_S(b,\theta)]^2+(\Gamma_{ph}/2)^2}M_S^{\rm eff}\,,
\end{align}
where $\Sigma_{ph}^<(\omega)=i\Gamma_{ph}n_F^S(\omega)$ and $\Sigma_{ph}^>(\omega)=-i\Gamma_{ph}(E_S)[1-n_F^S(\omega)]$, with $\Gamma_{ph}(E_S)$ being the finite width of the Shiba state induced by the photon-assisted excitations with the bulk. Note that the distribution function of the Shiba state reads:
\begin{equation}
n_F^S(\omega)=\frac{1}{e^{(\omega-b\cos{\theta})/T}+1}\,,
\end{equation} 
namely the quasi-static chemical potential is shifted by $b\cos{\theta}$ (the Shiba state is assumed to adiabatically follow the precession in the rotating frame). Next we can modify the Shiba Green's functions in order to account for the coupling to the STM tip. From the Dyson equation we obtain:
\begin{equation}
G_S^{\alpha}(\omega)=\frac{1}{1-g_S^{\alpha}(\omega)\Sigma_S^{\alpha}(\omega)}g_S^{\alpha}(\omega)\,,
\end{equation}
where $\alpha=r,a,<,>$, and
\begin{align}
\Sigma_S^{\alpha}(\omega)=\mathcal{T}^2\tau_zg^{\alpha}_T(\omega)\tau_z=\mathcal{T}^2\tau_zU(\theta)\tilde{g}_T^{\alpha}(\omega)U^\dagger(\theta)\tau_z\,,
\end{align}
is the Shiba state self-energy. Here,  
\begin{align}
\tilde{g}_T^\alpha(\omega) = \frac{1}{4}\sum_{\sigma,\tau=\pm1}\tilde{{\rm g}}_T^\alpha(\omega+\tau V+\sigma b)(1+\tau\tau_z)(1+\sigma\sigma_z)\,,
\end{align}
where
\begin{align}
\tilde{{\rm g}}_T^{r,a}(\omega)&=\mp 2\pi i\nu_0\,,\\
\tilde{{\rm g}}_T^{<}(\omega)&=2\pi i\nu_0n_F^T(\omega)\,,\\
\tilde{{\rm g}}_T^{>}(\omega)&=-2\pi i\nu_0[1-n_F^T(\omega)]\,.
\end{align}
We then obtain:
\begin{align}
\Sigma_S^{\alpha}(\omega)=(\mathcal{T}/2)^2\sum_{\sigma,\tau=\pm1}\tilde{{\rm g}}_T^\alpha(\omega-\tau V-\sigma b)(1+\tau\tau_z)[1+\sigma(\sigma_z\cos{\theta}+\sigma_x\sin{\theta})]\,.
\end{align}
That allows us to extract the renormalized Shiba Green's functions as follows:
\begin{align}
G_{S}^{a,r}(\omega)&=\frac{1}{\omega-E_S(b,\theta)\pm i\Gamma_{\rm tot}/2}M_S^{\rm eff}\,,\\
G_S^{<}(\omega)&=i\frac{n_F^S(\omega)\Gamma_{ph}+\sum_{\tau,\sigma}\Gamma_\tau n_{F,\tau,\sigma}^T(\omega)(1+\sigma\cos{\theta})}{[\omega-E_S(b,\theta)]^2+(\Gamma_{\rm tot}/2)^2}M_S^{\rm eff}\,,
\end{align}
while $G_S^{>}(\omega)$ is found by simply substituting all the Fermi-Dirac functions above as follows: $n_F\rightarrow-1+n_F(\omega)$. By utilizing the above expression, along with 
that for $g_{T}^\alpha(\omega)$, and inserting these into Eqs.~\eqref{Langreth}, we can easily obtain the expression for the charge current showed in Eq.~\eqref{tot_current}.  

\end{widetext}

\bibliography{Dynamical_Shiba_Final.bbl}

\end{document}